\documentclass[apj]{emulateapj}

\slugcomment{Accepted for publication in the Astronomical Journal}

\shorttitle{Evidence for the Evolution of Young Early-Type Galaxies in
  the GOODS/CDF-S Field} 
\shortauthors{Puzia, Mobasher, \& Goudfrooij}

\begin{document}

\title{Evidence for the Evolution of Young Early-Type Galaxies in the
  GOODS/CDF-S Field} 

\author{Thomas H. Puzia\altaffilmark{1,2,3}, Bahram
  Mobasher\altaffilmark{3,4}, and Paul Goudfrooij\altaffilmark{3}} 
\altaffiltext{1}{Herzberg Institute of Astrophysics, 5071 West Saanich
  Road, Victoria, BC, V9E 2E7, Canada} 
\altaffiltext{2}{Plaskett Fellow}   
\altaffiltext{3}{Space Telescope Science Institute, 3700 San Martin Drive,
    Baltimore, MD 21218}
\altaffiltext{4}{Affiliated with the Space Sciences Department 
of the European Space Agency}

\begin{abstract}
We have developed an efficient photometric technique for identifying young
early-type galaxy candidates using a combination of photometric redshifts,
spectral-type classification, and optical/near-infrared colors.~Applying
our technique to the GOODS HST/ACS and VLT/ISAAC data we have selected a
complete and homogeneous sample of young elliptical candidates among
early-type field galaxies. The distribution of structural parameters for
these candidates shows that their selection, which is based on early
spectral types, is fully consistent with early morphological types. We
investigate the evolution of their luminosities and colors as a function
of redshift and galaxy mass and find evidence for an increasing starburst
mass fraction in these young early-type galaxy candidates at higher
redshifts, which we interpret in terms of massive field galaxies
experiencing more massive/intense starbursts at higher
redshifts.~Moreover, we find indications for a systematically larger young
elliptical fraction among sub-$L^{\ast}/2$ early-type galaxies compared to
their brighter counterparts.~The total fraction among the field early-type
galaxies increases with redshift, irrespective of galaxy luminosity.~Our
results are most consistent with galaxy formation scenarios in which stars
in massive early-type field galaxies are assembled earlier than in their
low-mass counterparts.
\end{abstract}

\keywords{galaxies: K+A galaxies --- galaxies: formation and evolution}

\section{Introduction}
The last star-formation burst in galaxies defines their photometric
appearance and is an important diagnostic of galaxy formation and
evolution. The class of rejuvenated early-type galaxies includes so-called
`K+A' galaxies\footnote{Historically, these galaxies were named ''E+As''
because of their early-type morphology appearance on the very first
images. However, subsequent studies have shown that many of these galaxies
contain significant disk components, so that the ''E'' becomes unjustified
and is substituted by ''K'', owing to the spectral type of the underlying
old stellar population \citep{franx93}.} which are objects that show
spectroscopic signatures of old (K-type) and young (A-type) stellar
populations \citep{dressler83}.~The lack of emission lines in their
spectra indicates that star-formation processes abruptly ceased $\sim\!1$
Gyr ago, followed by a quiescent evolution into normal early-type galaxies
\citep[e.g.][]{couch87}.

Several scenarios have been suggested to explain this observed
rejuvenation phenomenon:~(1) galaxy-galaxy mergers
\citep[e.g.][]{zabludoff96}, (2) interactions of infalling galaxies with
the intracluster medium \citep[e.g.][]{gunn72, bothun86}, (3) galaxy
harassment \citep[e.g.][]{moore96, moore98}, (4) tidally induced star
formation \citep[e.g.][]{byrd90}, and (5) dusty starbursts
\citep[e.g.][]{poggianti99, poggianti00}. While tidal interaction and
galaxy harassment are important events in the overall context of galaxy
formation and evolution \citep[e.g,][]{bekki05}, they do not provide an
obvious explanation for the sudden halt of star formation without invoking
additional processes. Ram-pressure stripping and mergers are currently
seen as the most viable scenarios to describe the ignition and sudden
cessation of star formation in young early-type galaxies
\citep[e.g.][]{rose01, bekki03, shioya02, shioya04, quitnero04}.~Moreover,
recent radio observations appear to exclude a dusty starburst scenario in
general, as the measured radio fluxes are inconsistent with high star
formation rates (i.e. $\ga\!10-100\ M_{\odot}\ a^{-1}$) for virtually all
local post-starburst early-type galaxies \citep{miller01, goto04}.
However, \cite{smail99} does find evidence for faint radio fluxes in
spectroscopically confirmed post-starburst galaxies in distant dense
clusters at $z\approx0.4$.

K+A fractions are known to vary significantly with both redshift and
environmental density.~Several studies have revealed that the brightest
K+A's in nearby clusters are sub-$L^{*}$ systems \citep{caldwell99,
poggianti04}, while at intermediate redshifts ($z\ga0.8$) K+As are found
with $L\approx3 L^{*}$ \citep{dressler99, tran03}, suggesting an increase
of K+A cluster galaxy mass with redshift.~So far, the perhaps most
unbiased comparison between the K+A fractions in field and cluster
environments was done by \cite{tran03, tran04} who found a field K+A
fraction of $\sim\!3$\% and a factor $\sim\!4$ higher fraction in cluster
environment for the redshift range $0.3\!<\!z\!<\!1$.~\citeauthor{tran04}
further noted that the field K+A fraction shows strong fluctuations and is
sensitive to selection criteria. Furthermore, the relatively short
lifetime of the K+A signature \citep[$\sim\!1$ Gyr,][]{couch87, belloni95,
barger96} implies that these galaxies are rare.~This indicates the need
for wide-area surveys to identify statistically significant samples of
young early-type galaxies and determine their number fractions in both
field and cluster environments. If their selection is performed
consistently between datasets and galaxy formation models, their abundance
may be a sensitive diagnostic for star-formation activity in field and
cluster galaxies as a function of redshift, making these rejuvenated
objects useful tools to constrain hierarchical galaxy formation scenarios.

Keeping in mind that medium-resolution spectroscopy of all galaxies in
large-area surveys is very expensive in terms of telescope time, the aim
of this paper is to develop a photometric technique to identify young
early-type galaxy {\it candidates} (hereafter yE candidates) for
spectroscopic follow-up using a combination of photometric redshifts,
spectral-type classification, and optical/near-infrared color-color
diagrams.~Applying this technique to the GOODS/CDF-S dataset
\citep{giavalisco04}, we select a complete and homogeneous sample of yE
candidates in the redshift range $0\!<\!z\!\la\!1$.~Finally, we
investigate correlations between the colors of yE candidate galaxies with
redshift and galaxy mass, which, if confirmed by spectroscopy, may put
strong constraints on galaxy formation models.~We use the standard
$\Lambda$CDM cosmology parameters $\Omega_{M}=0.3$,
$\Omega_{\Lambda}=0.7$, and $H_{0}=70$ km s$^{-1}$ Mpc$^{-1}$ throughout
this work.

\section{GOODS/CDF-S Data}

One of the two fields covered by the Great Observatories Origins Deep
Survey  (GOODS) Treasury Program is the {\it Chandra Deep Field South}
(CDF-S) with optical HST/ACS and ground-based VLT/ISAAC near-infrared
photometry \citep{giavalisco04}.~These data span the wavelength range from
near-UV to the near-infrared and provide ACS imaging with very high
spatial resolution (0.03\arcsec) for $\sim\!86000$ galaxies down to
$V\approx27.5$.~Object detection and photometry was performed with
SExtractor \citep{bertin96} after the images in all passbands were
convolved to a common FWHM$\approx\!0.45$\arcsec.~The final magnitudes are
the MAG\_AUTO SExtractor magnitudes, which measure the total light.

Using the multi-waveband data available for galaxies in this field,
photometric redshifts were measured following the technique presented in
\cite{mobasher07}. Briefly, this approach compares the template Spectral
Energy Distributions (SEDs) for different spectral types of galaxies,
shifted in redshift space, with the observed SED for galaxies detected in
GOODS (see also Sect.~\ref{spectypes}). $\chi^2$ fits are performed and
the redshift and spectral type corresponding to the minimum $\chi^2$ value
are associated to those for the galaxy in question. The template spectral
types consist of elliptical, Sbc, Scd, and Im-type galaxies from
\cite{coleman80} and starburst templates from \cite{kinney96}.
Bayesian priors based on observed galaxy luminosity functions are used
\citep[as detailed in][]{mobasher07}. Comparison with galaxies with
available spectroscopic redshifts yields a photometric redshift accuracy
of $\sigma(\Delta z)\!=\!0.03$ where $\Delta z=(z_{\rm phot}-z_{\rm
spec})/(1+z_{\rm spec})$.

In order to check the accuracy of our spectral types, we compare these
with the morphology of galaxies measured from the GOODS-S ACS images. The
morphologies are estimated using the concentration and asymmetry
parameters \citep{conselice97, abraham96}, measured for individual
galaxies  (B. Mobasher- private communication). We found a very good
correlation between the morphology parameters and spectral types,
particularly for elliptical galaxies. Therefore, we have reliable
identification of elliptical galaxies, based on their spectral types,
which constitutes the main selection criterion for the present study.

We use the information on redshifts and spectral types to compute
$K$-corrections for all galaxies in our sample. Throughout the
rest of this work we use AB magnitudes \citep{oke74}\footnote{AB
magnitudes are defined by $m_{\rm AB}=-2.5\log f_{\nu}-48.6$, where
$f_{\nu}$ is in ergs s$^{-1}$ cm$^{-2}$ Hz$^{-1}$. Alternatively, one can
write $m_{\rm AB}=-2.5\log f_{\nu}+8.9$, where $f_{\nu}$ is the flux in
Jy.}, and correct for Galactic foreground reddening $E_{B-V}=0.008$
derived from the \cite{schlegel98} maps. We select galaxies brighter than
$M_{K}\!=\!-19.0$ mag from the initial GOODS/CDF-S sample which have a
mean photometric redshift accuracy of $\Delta z\leq0.15$ corresponding to
a 95\% confidence interval. This photometric redshift uncertainty limit is
introduced in order to keep the average color uncertainty below 0.15 mag
due to k-corrections (see also Sect.~\ref{ln:ssp}).~The final sample is
volume-selected and complete to $z\approx1$. We exclude objects where the
photometry was compromised by saturated pixels, corrupted or truncated
isophotal apertures, and where the extraction algorithm experienced
problems during deblending or the photometric errors were larger than
$\Delta m=0.15$ mag in any filter. The final catalog includes objects
with HST/ACS $BViz$ and VLT/ISAAC $JHK_s$ magnitudes\footnote{In the
following we shall refer to the $K_s$ filter as the $K$ band.}.


\section{Stellar Population Models}
\label{ln:ssp}

\subsection{Diagnostic Colors}
Based on the \cite{bruzual03} simple stellar population (SSP) models and
their 2003 version of the GALAXEV stellar population synthesis code, we
compute optical/near-infrared photometric colors for composite stellar
populations (CSP).~We use the Salpeter IMF with limits at $0.1$ and $100
M_{\odot}$.~By comparison with the control sample of spectroscopically
confirmed K+As, we identify the best diagnostic combination
optical/near-infrared colors to select yE candidates.

Given the availability of many filters in our GOODS dataset
($BVizJHK_{s}$) we investigate several filter combinations that may most
efficiently reduce the age-metal\-licity-extinction degeneracy in the
color-color plane. Due to the presence of a hotter $\sim1$ Gyr young
stellar population, the $B$-band flux of yE galaxies is expected to be
enhanced relative to older ($\gg1$ Gyr) stellar populations.~For slightly
younger stellar populations with ages $\sim\!0.4\!-\!0.8$ Gyr the
Asymptotic Giant Branch (AGB) is densely populated with relatively cool
stars \citep{renzini81} causing a stronger output in the near-infrared
$K$-band \citep{persson83}.~Since a small population of luminous
thermally-pulsing AGB stars (TP-AGB) can significantly contribute to the
integrated light of galaxies, the consideration of such short-lived
evolutionary phases in SSP models is essential \citep{maraston98,
mouhcine03}.~Past this so-called AGB-phase transition, near-infrared light
is primarily sensitive to the mean temperature of the red giant branch,
which is mainly driven by the luminosity-weighted mean metallicity of
composite stellar populations.

\begin{figure*}[!t]
\centering 
\includegraphics[bb=0 100 400 400, width=12cm]{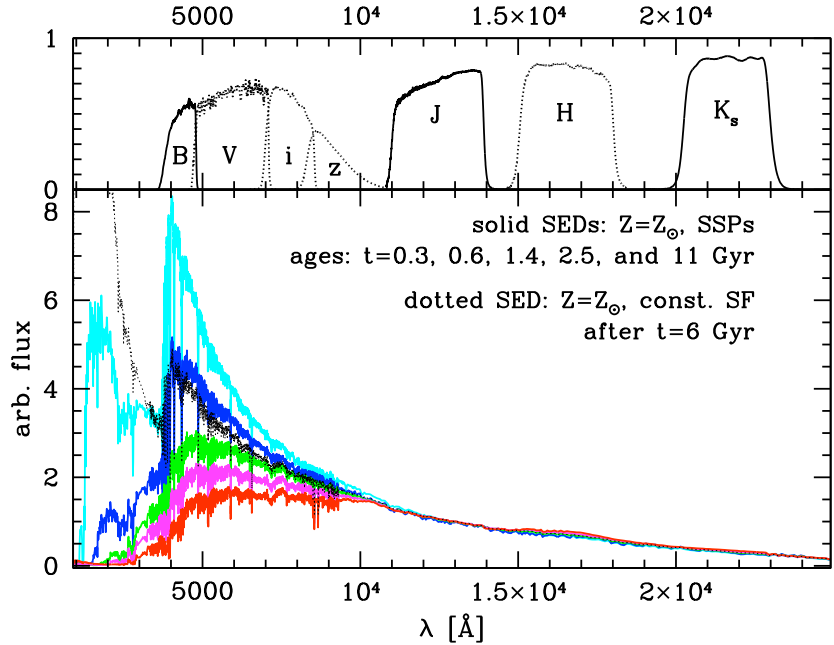}
 \caption{Comparison of restframe solar-metallicity SEDs for 
 single-burst stellar populations with ages $t=290$ Myr, 640 Myr, 1.4 Gyr,
2.5 Gyr, and 11 Gyr.~A dotted line illustrates the SED of a
solar-metallicity stellar population with a constant star formation rate
after 6 Gyr.~All SEDs were taken from \cite{bruzual03}.~Total filter 
throughputs for ACS and the ISAAC camera, including mirror reflectivity, 
window transmission, and detector quantum efficiency, are shown at 
the top of the panel.}
\label{ps:sedsolar}
\end{figure*}

In Figure~\ref{ps:sedsolar} we show the effective filter throughputs for
HST/ACS and the VLT/ISAAC instrument together with spectral energy distributions
(SEDs) for single-burst stellar populations for ages $t=290$ Myr, 640
Myr, 1.4 Gyr, 2.5 Gyr, and 11 Gyr at solar metallicity. For illustration
purposes all SEDs were normalized to the flux at $1.3\ \mu m$. It is clear
from the figure that the $B$ band contains the most age-sensitive flux and
that the $B\!-\!J$ color is a very good age indicator.~It is instructive
to note that the Lyman break is a sensitive indicator to identify
constantly star-forming or very young post-starburst galaxies
(i.e.~$t\la0.5$ Gyr, see dotted SED in Fig.~\ref{ps:sedsolar}).~However,
the Balmer break is the more sensitive proxy to identify post-starburst
galaxies with intermediate ages (i.e.~$t\ga0.5$ Gyr), such as K+A
galaxies.~To find a good metallicity indicator we need to move to the
near-infrared, past the $z$-band, where the continuum slope shows little
impact of varying age.~Here, the $H$-band continuum suffers from
variations due to strong water vapor absorption, which is sensitive to the
mean temperature of the stellar population
(i.e.~age-sensitivity).~However, the relative continuum fluxes in $J$ and
$K$ show little variation with age, and we define $J\!-\!K$ as our best
metallicity indicator which exhibits only modest age sensitivity during
the AGB-phase transition \citep[see also][]{ferraro00}.

The best choice of the most age-sensitive and the most
metallicity-sensitive color depends also on data quality.~For reference,
the average color uncertainty introduced by the photometric redshift error
is $\la0.15$ mag in both $B\!-\!J$ and $J\!-\!K$ for all galaxy types at
$z\la1$. Given the mean uncertainty of all color combinations of our data,
the most age-sensitive color is $B\!-\!J$ and the most
metallicity-sensitive color is $J\!-\!K$, which is virtually independent
of dust reddening.~This $BJK$ combination optimizes the selection of
post-starburst intermediate-age stellar populations at redshifts
$z\!\la\!1$.~For higher redshifts, the age-metallicity degeneracy dilutes
the age/metallicity sensitivity of this color combination as the restframe
flux at $\lambda\!<\!1\ \mu m$ moves into the $J$ band, and a
different filter combination is required.

A comparable photometric selection technique, although specifically
targeting star-forming galaxies at higher redshifts, was described by
\cite{daddi04}.~Similar techniques to detect intermediate-age globular
clusters (i.e.~single-burst stellar populations) in nearby galaxies have
been recently used by \cite{goudfrooij01}, \cite{puzia02}, and
\cite{hempel04}.

\begin{figure}[!b]
\centering 
\includegraphics[bb=125 100 400 400, width=5cm]{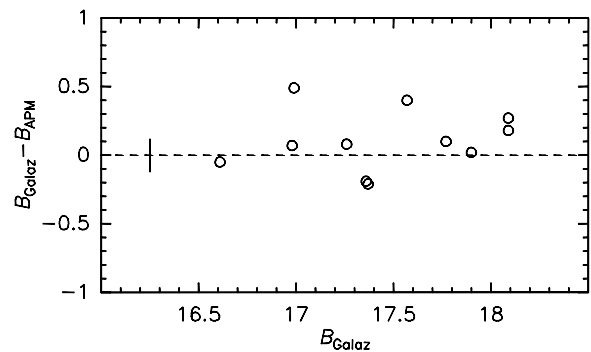}
 \caption{Comparison of $B$ magnitudes for spectroscopically confirmed K+A galaxies
taken from \cite{galaz00} with those taken from the APM catalog.~A vertical solid line 
shows the mean formal uncertainty.}
\label{ps:compphot}
\end{figure}

\begin{figure*}[!ht]
\centering
 \includegraphics[bb=0 0 400 400, width=8cm]{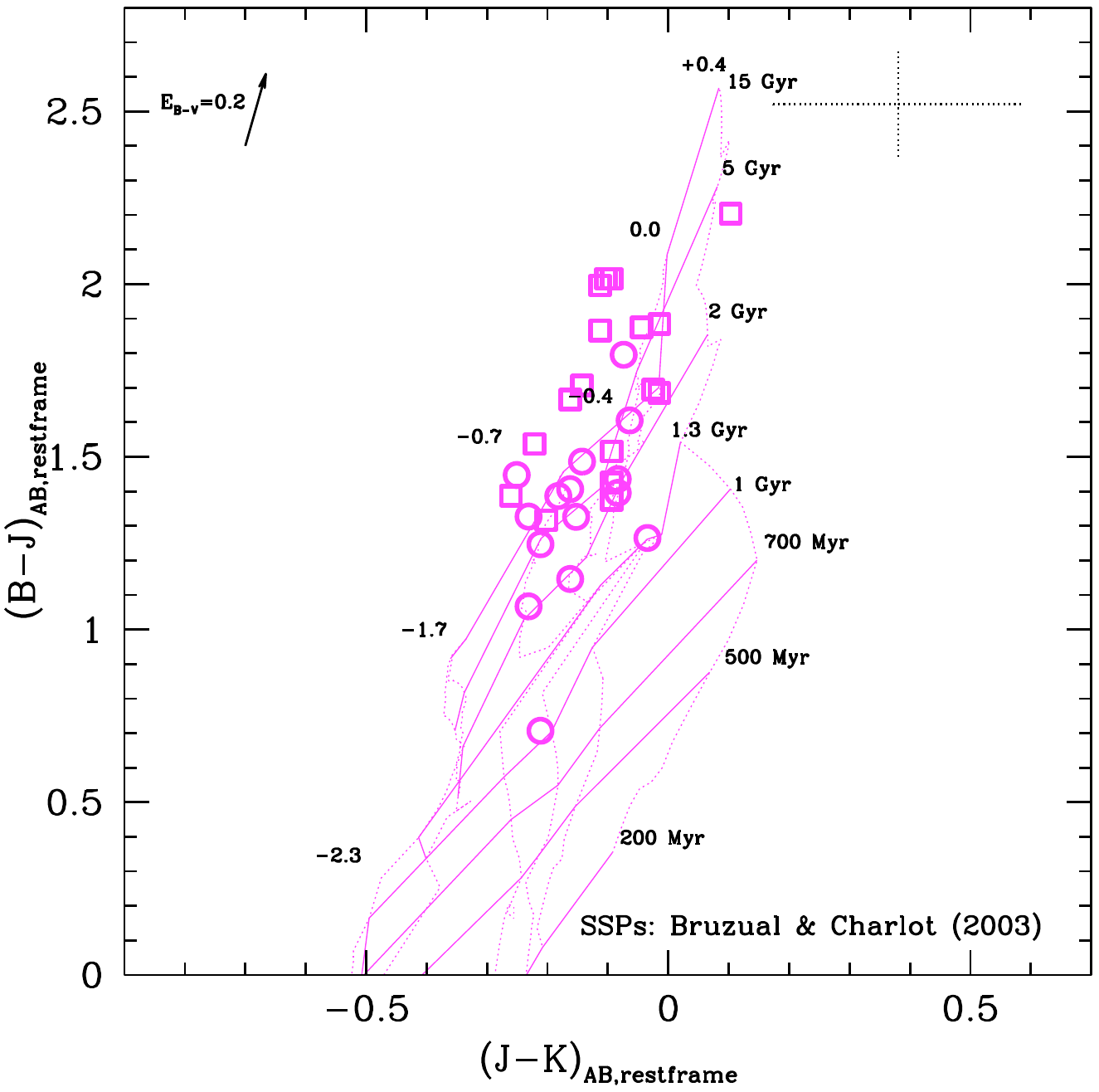}
 \includegraphics[bb=0 0 400 400, width=8cm]{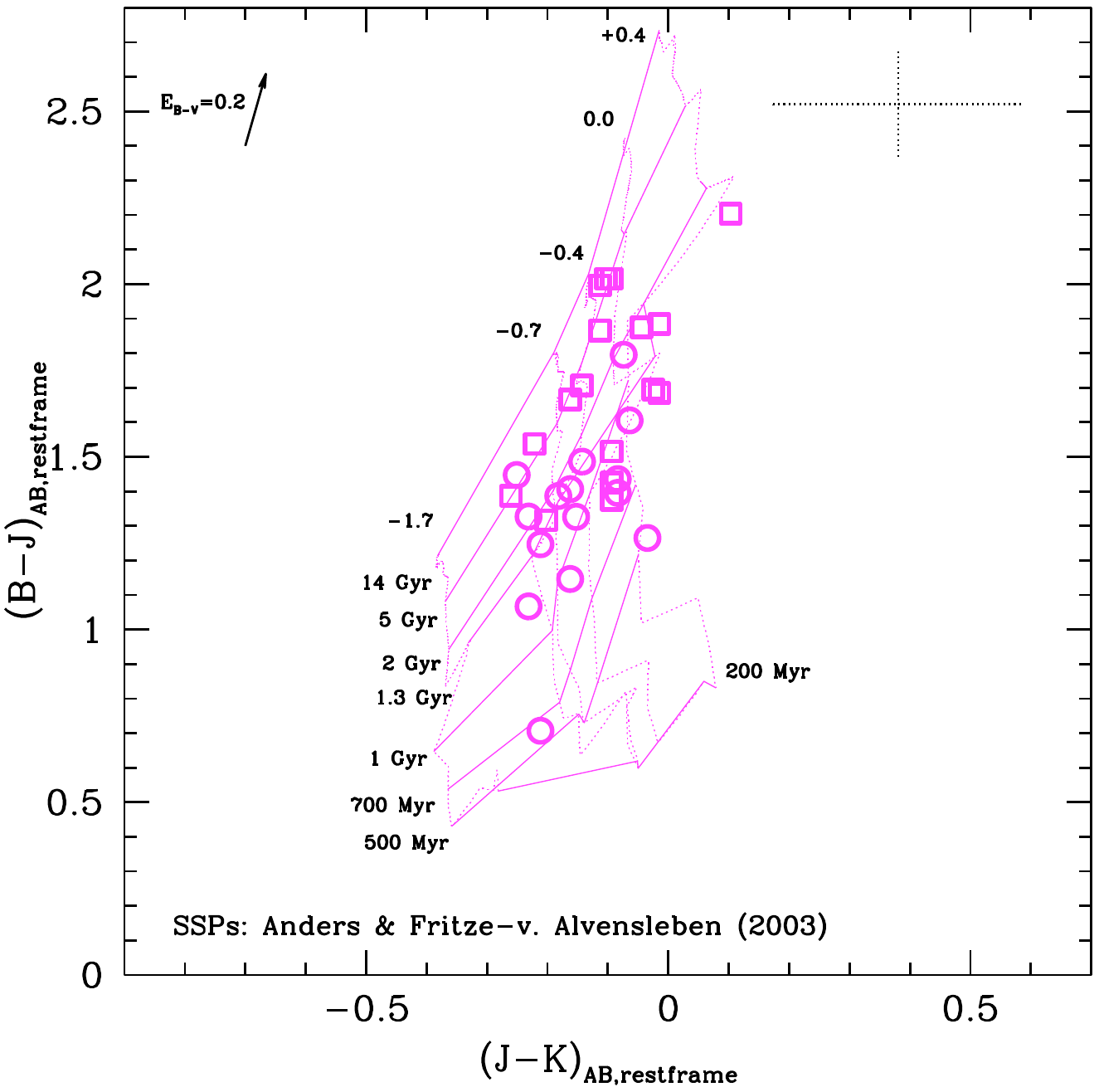}
 \caption{{\it Left panel}: Comparison of $(B\!-\!J)_{\rm AB}$ 
 vs.~$(J\!-\!K)_{\rm AB}$ restframe colors of spectroscopically confirmed
K+A galaxies ({\it open symbols}) with SSP models of \cite{bruzual03}. Open
squares mark cluster K+A galaxies, open circles are field K+As.~Isochrones 
({\it solid lines}) are plotted for ages $t=200$,
500, 700 Myr, and 1, 1.3, 2, 5, and 15 Gyr.~Iso-metallicity lines ({\it dotted
lines}) are shown for the metallicities [Z/H]~$=-2.3, -1.7, -0.7, -0.4,
0.0$, and $+0.4$ dex.~{\it Right panel}:~Same data, but this time with SSP
models of \cite{anders03}. Here isochrones ({\it solid lines}) were plotted for
the ages $t=200$, 500, and 700 Myr, and 1, 1.3, 2, 5, and 14 Gyr.
Iso-metallicity tracks ({\it dotted lines}) are indicated for [Z/H]~$=-1.7,
-0.7, -0.4, 0.0$,  and $+0.4$ dex. Each panel has the extinction vector
for E$_{B-V}=0.2$ mag indicated in the upper left corner. Mean uncertainties 
of the \cite{galaz00} data are shown in the upper right corner.}
\label{ps:sspcomp}
\end{figure*}

\subsection{Spectroscopically confirmed K+As}
To test our photometric selection technique we use the K+A galaxy sample
of \cite{galaz00}.~This data set contains spectroscopically confirmed K+As
in nearby ($z\approx0.05$) and distant ($z\approx0.3$) clusters and in the
nearby field ($z\approx0.1$).~All galaxies have been classified as K+As
\citep{couch87, franx93, caldwell97, zabludoff96}, without any indications
of emission lines.~The near-infrared photometry of the
\citeauthor{galaz00} study was performed with the Las Campanas NICMOS3
Hg:Cd:Te detector in the CIT filter system \citep{persson98}. We
compute the transformations to the \citep[][hereafter BB]{bb88} standard system
using \citet{bruzual03} SEDs with ages from 0.2
Gyr to 15 Gyr and [Fe/H] values from $-2.25$ to $+0.4$ dex, and find the
following (BB)--(CIT) transformations
\begin{eqnarray*}
J_{\rm BB} &=& J_{\rm CIT} - 0.0045 + 0.0102\, (J-K)_{\rm CIT} \\
(J-K)_{\rm BB} &=& 0.0147 + 0.9835\, (J-K)_{\rm CIT} \\
\end{eqnarray*}
with RMS scatter of 0.0009 mag and 0.007 mag, respectively.~Galaz'
Table~2 lists Johnson B magnitudes as reported in NED\footnote{the
NASA/IPAC Extragalactic Database, http://nedwww.ipac.caltech.edu} at the
time, likely coming from various original sources. We compare these $B$
magnitudes with those in the APM
catalog\footnote{http://www.ast.cam.ac.uk/\~{}apmcat} \citep{maddox90} and
show the comparison plot in Figure~\ref{ps:compphot}.~We find significant
offsets for the optical photometry between the \citeauthor{galaz00} and
the APM catalog.~Since the magnitudes in the APM catalog were derived in
an internally consistent way, we adopt the APM $B$ magnitudes for further
analysis in this paper.~The mean uncertainties in $BJK$ colors of the
comparison sample are 0.12, 0.10, and 0.18 mag, respectively.~All optical
and near-infrared magnitudes from the GOODS catalog and the comparison
sample represent total magnitudes.

\subsection{Confirmed K+As in the $BJK$ color-color plane}
In Figure~\ref{ps:sspcomp} we compare $(B\!-\!J)_{\rm AB}$
vs.~$(J\!-\!K)_{\rm AB}$ restframe colors of spectroscopically confirmed
K+A galaxies with SSP model predictions from two different groups:
\citeauthor{bruzual03} (\citeyear{bruzual03}, hereafter BC03) and
\citeauthor{anders03} (\citeyear{anders03}, hereafter AF03).~The BC03
predictions were calculated in the AB system with the same filter
transmission functions as the GOODS/CDF-S data. For the AF03 models, which
are tabulated in Vega-based magnitudes using the BB filter system, 
we follow the Vega-AB transformations: $K_{\rm AB}=1.891+K_{\rm Vega}$, $J_{\rm
AB}=0.91+J_{\rm Vega}$, and $B_{\rm AB}=-0.105+B_{\rm Vega}$.~As expected,
the location of confirmed K+As in the color-color plane corresponds to old
to intermediate ages. We find that the absolute age calibration differs
between the two models.~The BC03 models predict (luminosity-weighted) ages
between $\sim\!15$ and $\sim\!2$ Gyr for most K+As, while the models of
AF03 predict younger ages in the range $\sim\!1\!-\!5$ Gyr.~The
metallicity scales on the other hand are very similar in both models. The
majority of confirmed K+A galaxies have luminosity-weighted metallicities
between [Z/H]~$\approx-0.7$ and $0.0$ dex, with few outliers at
super-solar metallicities, which might be reddened objects.

\subsection{Composite Stellar Populations in the BJK Color-Color Plane}
Several groups compared the mean colors of K+A galaxies using optical and
near-infrared photometry with population synthesis models
\citep[e.g.][]{newberry90, belloni95, barger96, shioya02, balogh05}.~In
this work we combine spectral-type fitting with model predictions and
investigate the influence of different metallicities and mass fractions of
the starburst population on colors and luminosities of post-starburst
galaxies in the optical/near-infrared color-color plane and in
color-magnitude diagrams.

\begin{figure*}[!ht]
\centering
 \includegraphics[bb=0 0 400 400, width=8cm]{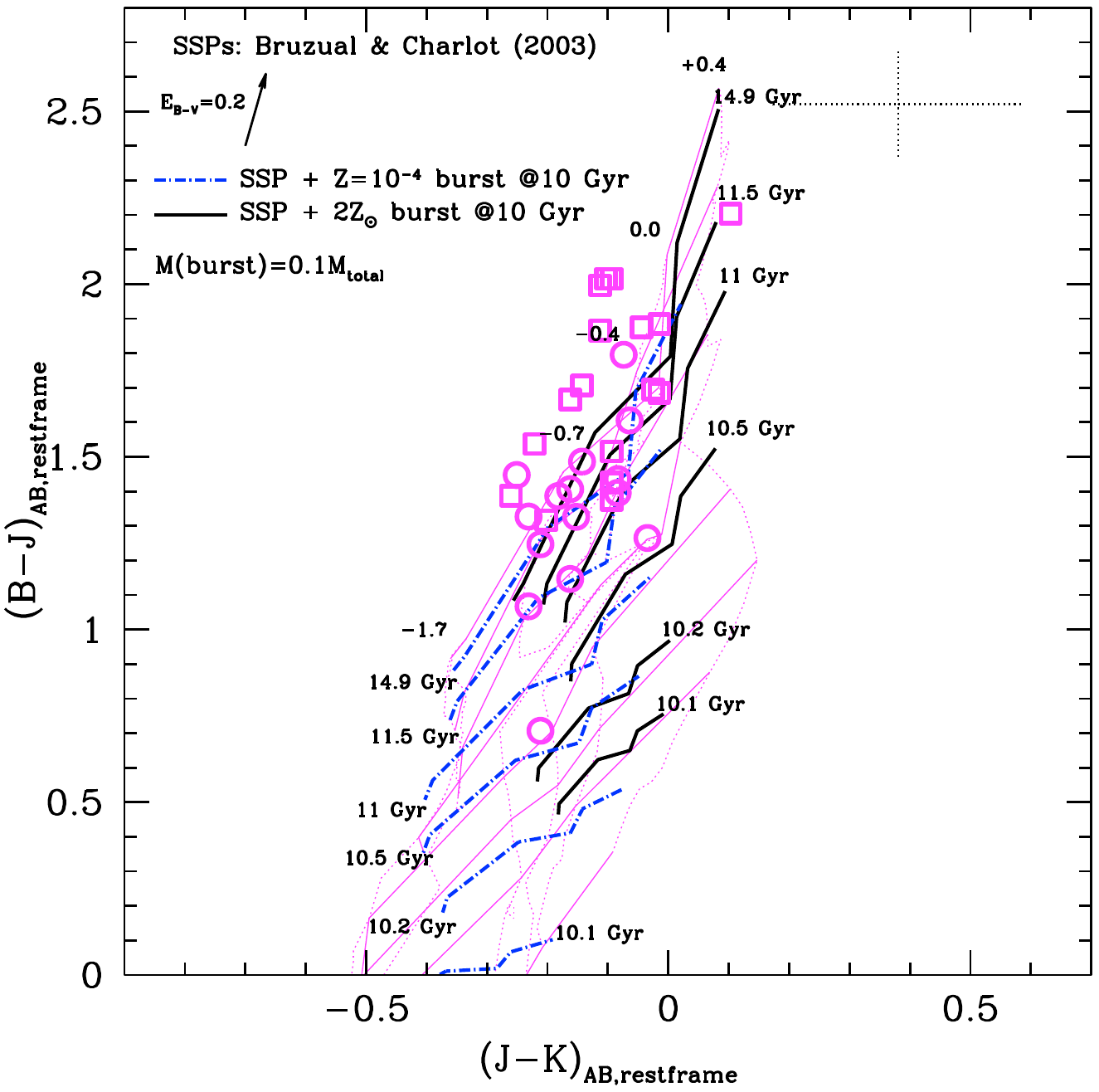}
 \includegraphics[bb=0 0 400 400, width=8cm]{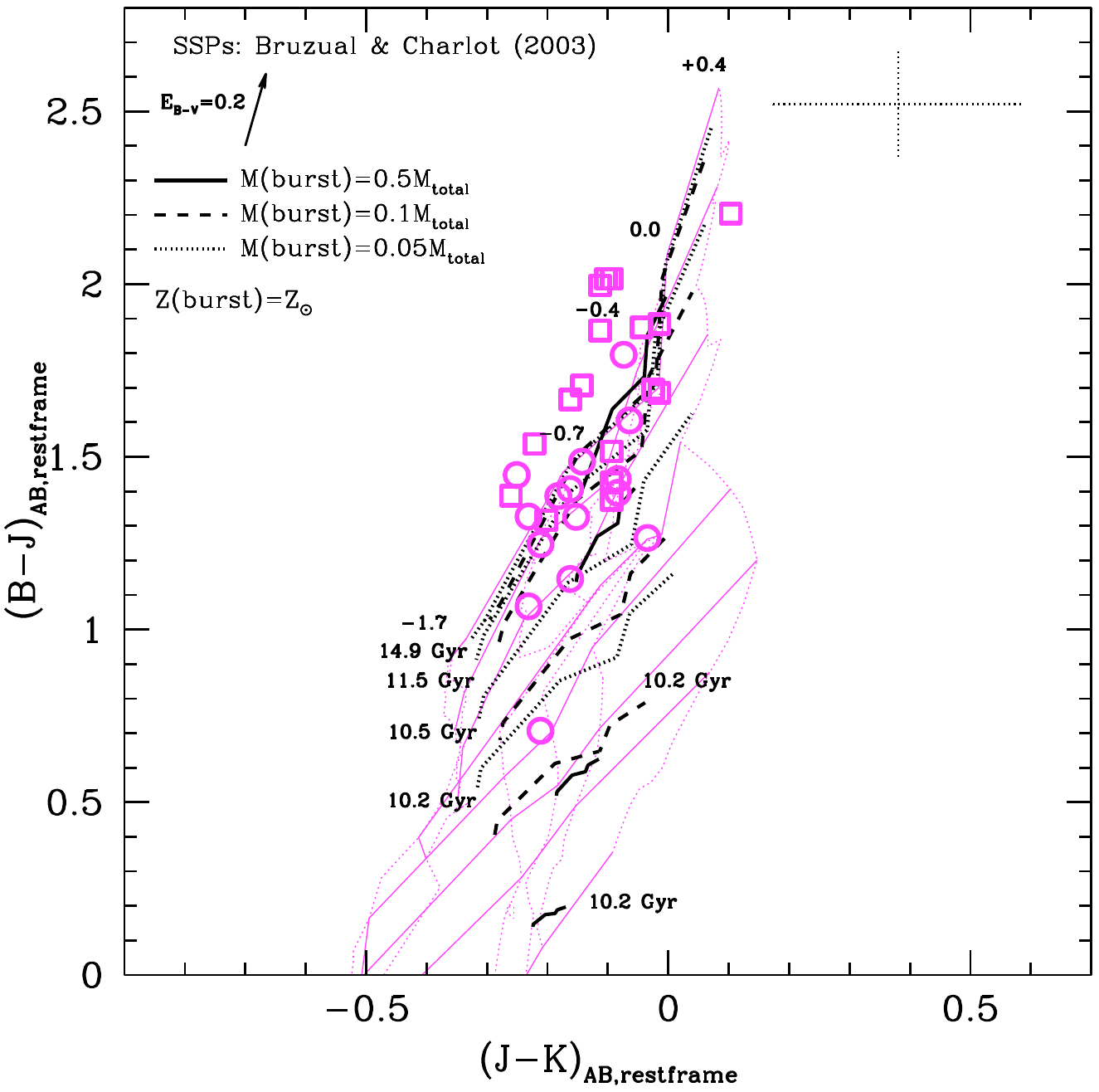}
 \caption{{\it Left panel}: Evolution of $(B\!-\!J)_{\rm AB}$
vs.~$(J\!-\!K)_{\rm AB}$ restframe colors for an SSP with a starburst that
ignites after 10 Gyr of passive evolution and decays for another 5
Gyr.~Two set of isochrones with ages $t=10.1, 10.2, 10.5, 11, 11.5$, and
14.9 Gyr are shown for the metallicities [Z/H]~$=0.4$ ({\it solid lines})
and $-2.3$ dex ({\it dot-dashed lines}) of the starburst stellar
population.~The starburst strength amounts to 10\% of the total final stellar
mass.~{\it Right panel}:
Evolution of colors for different starburst strengths/masses for a solar
metallicity burst.~Ischrones for ages $t=10.2, 10.5, 11.5$, and 14.9 Gyr
are plotted with different line types ({\it solid, dashed,} and {\it
dotted lines}) indicating the dependence on starburst mass $M_{\rm
burst}=0.05, 0.1$, and 0.5 $M_{\rm total}$. For reference we overplotted
in both panels the SSP model grid from the left panel of
Figure~\ref{ps:sspcomp}.~All predictions were computed using the SSP
models of \cite{bruzual03}. Spectroscopically confirmed K+A galaxies are 
as in Figure~\ref{ps:sspcomp}.}
\label{ps:sspburst}
\end{figure*}

Figure~\ref{ps:sspburst} shows the evolution of BC03 isochrones where a
starburst is ignited after 10 Gyr of passive SSP evolution.~Assuming a
minor merger, we assign 10\% of the final stellar mass to the starburst
population. We adopt two metallicities Z~$=0.04$ (twice solar metallicity)
and Z~$=10^{-4}$ ([Z/H]~$=-2.3$ dex) for the starburst stellar population,
which are superposed on top of the full range of metallicities
([Z/H]~$=-2.3$ to $+0.4$ dex.) of the underlying, older stellar
population.~Note the significant difference in $B\!-\!J$ colors during
the first $\sim\!1.5$ Gyr after the starburst.~If determined relative to
passively evolving galaxies, this color offset is a sensitive
post-starburst indicator and will be discussed below in more detail.

In the right panel of Figure~\ref{ps:sspburst} we show the influence of
changing starburst mass fraction on the post-starburst galaxy colors.~This
time we choose solar metallicity for the starburst with mass
fractions are 5\%, 10\% (minor merger), and 50\% (major merger) of the
final total stellar mass. The Figure illustrates that the higher the
starburst mass fraction the bluer the $B\!-\!J$ color is during the
post-starburst phase, in agreement with previous studies
\citep[e.g.][]{barger96}.~In addition, there is a mass
fraction-metallicity degeneracy of post-starburst near-infrared colors.
Metal-rich major starbursts can exhibit the same $J\!-\!K$ colors as
metal-poor minor starbursts.~This degeneracy is most prominent for
intermediate metallicities in the range $-0.7\la$~[Z/H]~$\la0$.~For more
extreme metallicities the burst strength can be constrained with
increasing confidence.~In other words, yE galaxies with starburst mass
fractions $\la 10$\% and $J\!-\!K\ga0$ are likely to have experienced a
recent starburst with super-solar metallicity.~The same mass-fraction
limit and $J\!-\!K\la-0.3$ mag indicate sub-solar starburst metallicities.

In summary, $B\!-\!J$ is mainly degenerate in age and starburst mass
fraction, $J\!-\!K$ is mainly degenerate in metallicity and starburst mass
fraction.~We conclude that the $B\!-\!J$ vs. $J\!-\!K$ color-color plane
can be used to identify post-starburst galaxies.~But only extreme
metallicities of the post-starburst stellar population can be narrowed
down robustly.~The color-color plane is not suitable to derive starburst
mass fractions for intermediate metallicities without any other
information.~However, these degeneracies can be partly lifted by including
restframe $K$-band luminosities.~Our findings are relatively independent
of the choice of SSP models, as the systematics in the AF03 models give
the same results.


\section{Selection of {\footnotesize y}E Candidates}
\subsection{Spectral Types} 
\label{spectypes}
\begin{figure}[!t]
\centering 
\includegraphics[bb=0 0 400 400, width=8cm]{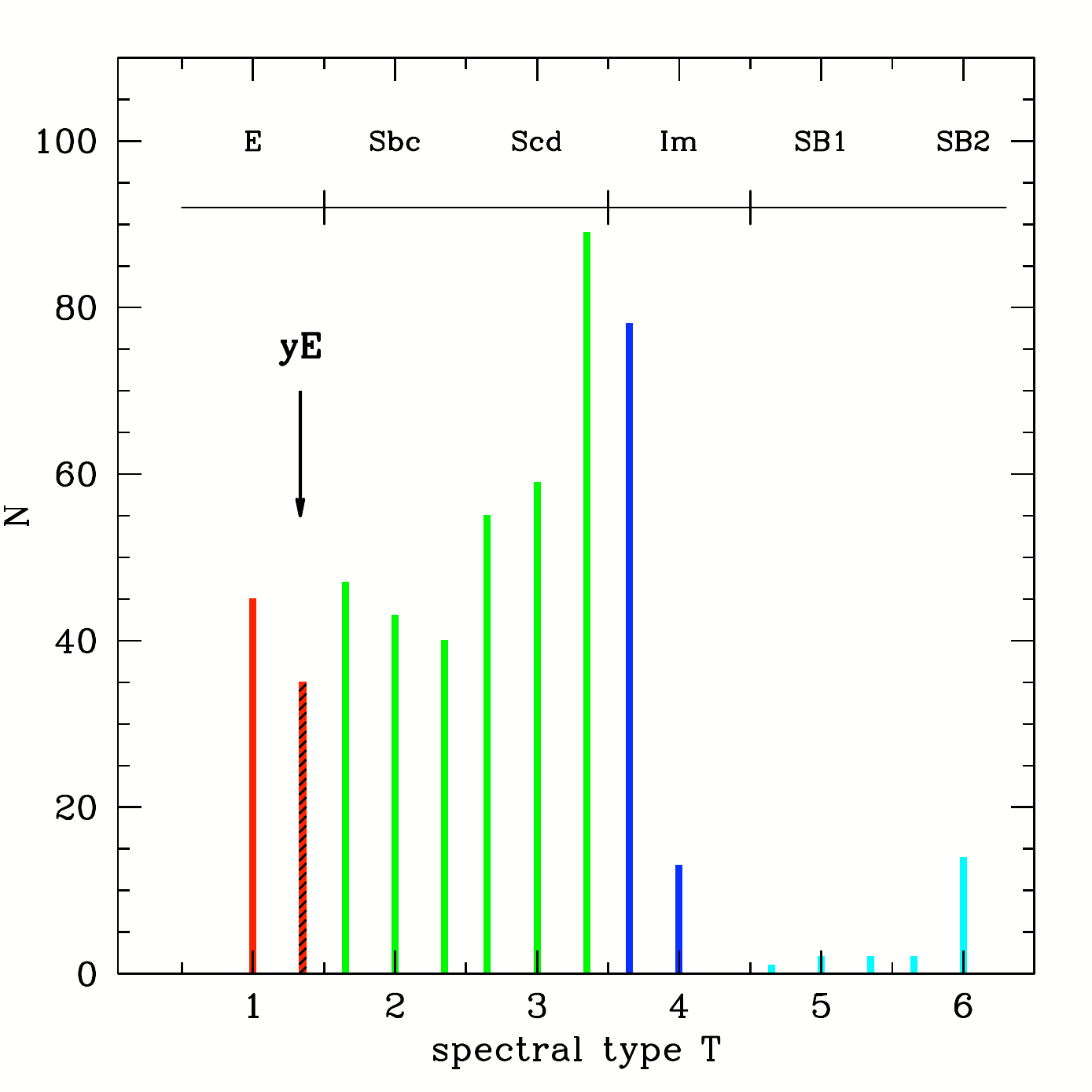}
 \caption{Distribution of spectral types $T$ that were assigned by the photometric
redshift routine for all objects in our selected GOODS/CDF-S sample
\citep[for details on the photometric redshift fitting routine we refer the reader
to][]{mobasher04, mobasher07}.~The locations of individual spectral types are labeled.~A 
sub-population of several objects have a slightly earlier spectral type ($T\!=\!1\onethird$)
than passively evolving galaxies ($T\!=\!1$).~These are our yE candidates
and are marked in the plot with a hatched bin.}
\label{ps:spectype}
\end{figure}

\begin{figure}[!th]
\centering 
\includegraphics[bb=0 0 400 400, width=8cm]{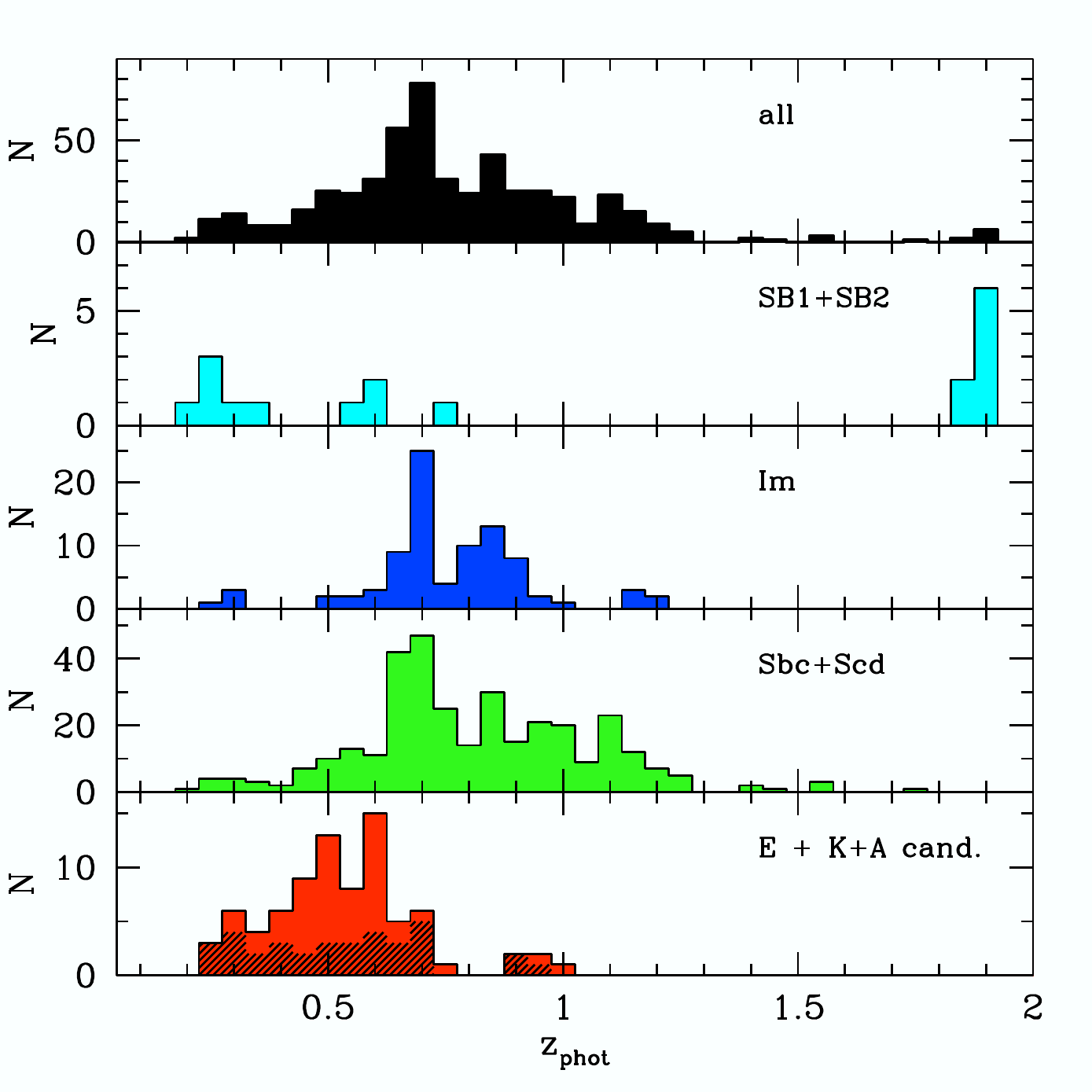}
 \caption{Distribution of photometric redshifts for all selected objects 
 in our GOODS/CDF-S photometric dataset \citep[for details 
 on the photometric redshift routine see][]{mobasher04}. The parametrization 
 by spectral-type is identical to the one in Figure~\ref{ps:spectype}.}
\label{ps:redshift}
\end{figure}

In the following we develop a robust method to identify yE galaxy
candidates from the GOODS/CDF-S photometric catalog using a combination of
photometric redshifts, spectral-type classification, and
optical/near-infrared colors.~The photometric redshifts and spectral types
are measured by using template SEDs, shifting them in redshift steps and
performing $\chi^2$ fits to the observed SEDs at each step. The templates
corresponding to the minimum $\chi^2$ values are selected, with its
spectral type and redshift assigned to the galaxy in question. We use the
observed templates for elliptical, Sbc, Scd and Im-type galaxies from
\cite{coleman80} and the starburst SEDs from \cite{kinney96}. In order to
increase the resolution in spectral types, we also divide the interval
between adjacent spectral classes into two intermediate bins. For
instance, between the two discrete spectral types for ellipticals
($T\!=\!1$) and Sbc-type spectra ($T\!=\!2$) we include two discrete
sub-categories ($T=1\onethird, 1 \frac{2}{3}$) that are linear mixtures
between the two boundary types. This allows a more continuous spectral
classification across the Hubble types while not causing degeneracy. The
choice of steps of 1/3 unit was made as a compromise between the goal to
increase the resolution of spectral types on one side and the goal to
avoid creating degeneracies in the phot-z routine on the other. Details of
the photometric redshift and spectral-type measurement are presented in
\cite{mobasher04, mobasher07}.~We plot the distribution of the discrete
spectral types in Figure~\ref{ps:spectype} and define all objects that
were assigned slightly earlier spectral types ($T\!=\!1\onethird$) than
passively evolving galaxies ($T\!=\!1$) as yE candidates, i.e.~objects
which are bluer than normal ellipticals but redder than normal early-type
spirals.~Their location in Figure~\ref{ps:spectype} is marked by a hatched
histogram.~Because later galaxy types can exhibit similar $BJK$ colors as
genuine yE galaxies, we use the spectral-type information to exclude
late-type star-forming galaxies and to include only galaxy candidates with
early spectral types in the following analysis.~The distribution of
photometric redshifts of all objects is shown in Figure~\ref{ps:redshift}.
Most of our yE candidates have redshifts between $z\approx0.2$ and $0.8$,
with few objects at slightly higher redshifts $z\approx1.0$.

\begin{figure*}[!t]
\centering 
\includegraphics[bb=0 0 400 400, width=5.69cm]{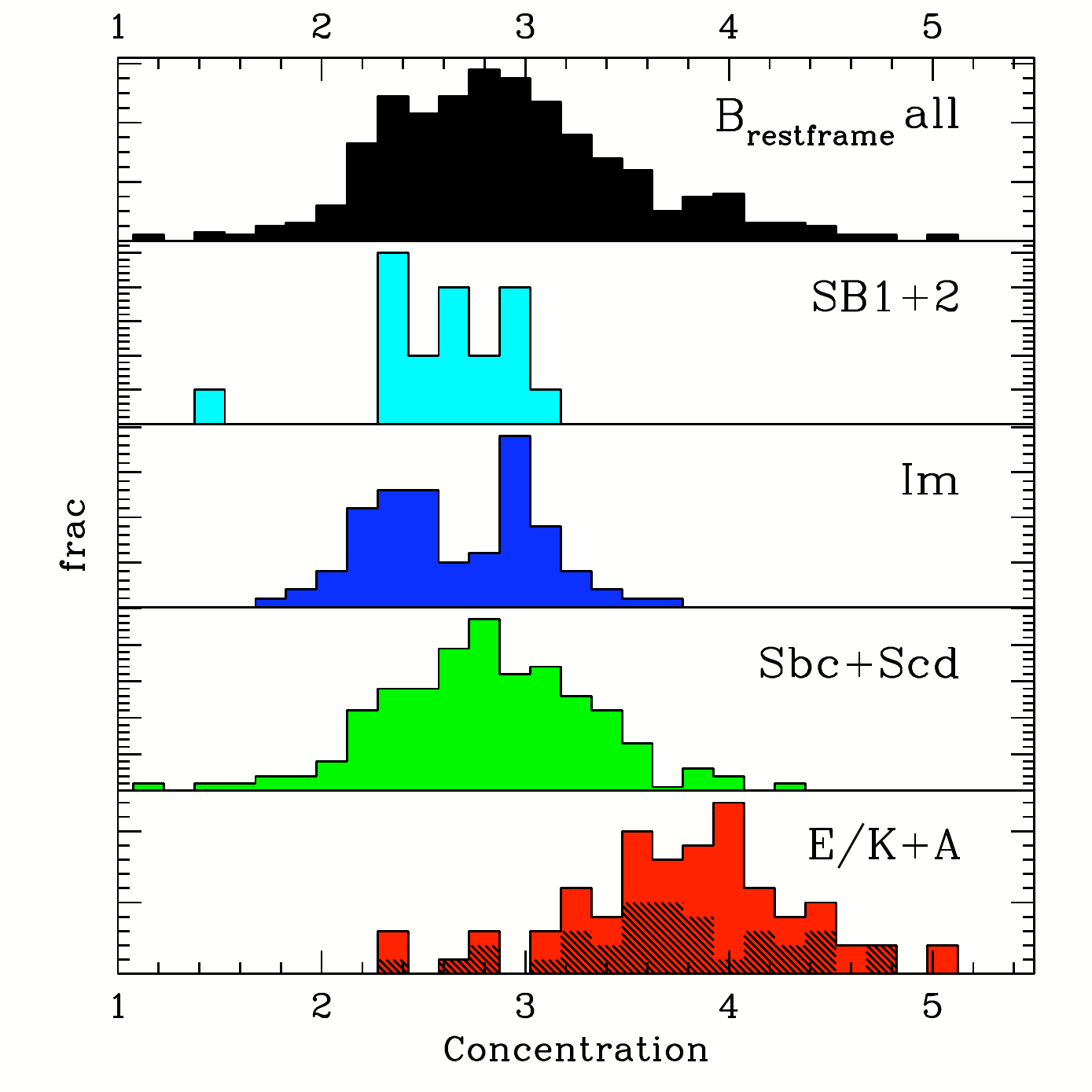}
\includegraphics[bb=0 0 400 400, width=5.69cm]{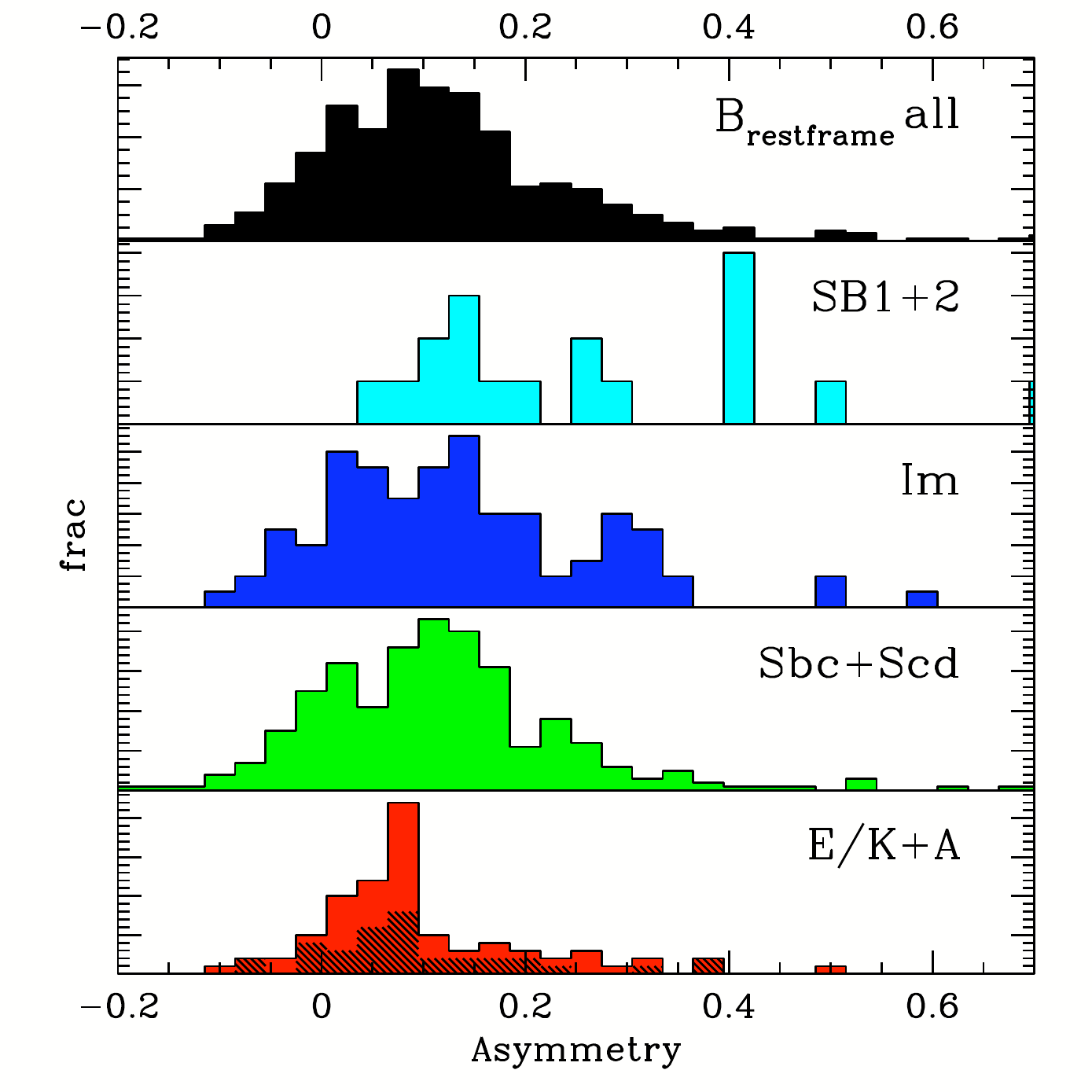}
\includegraphics[bb=0 0 400 400, width=5.69cm]{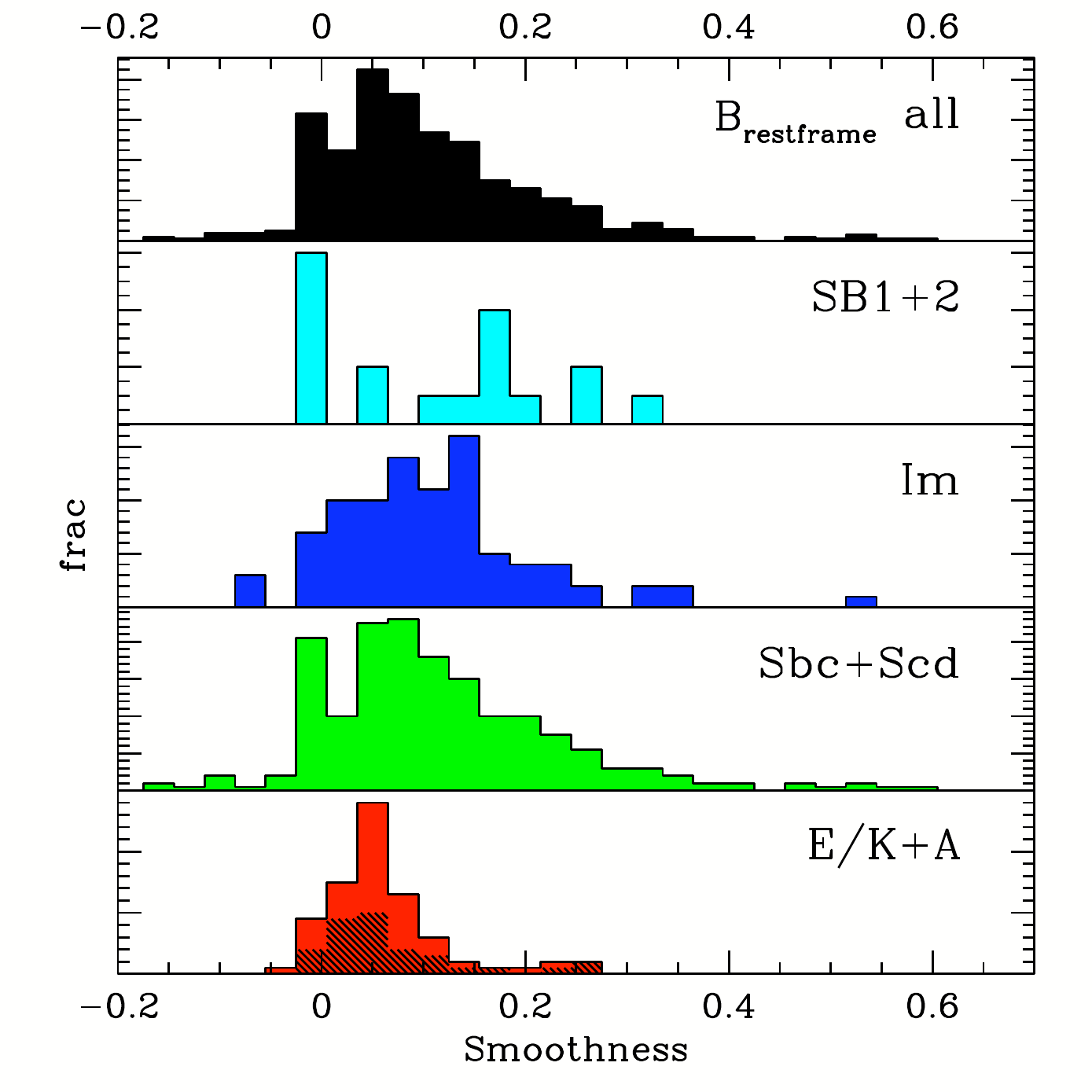}
 \caption{Distribution of restframe $B$-band CAS parameters \citep[concentration, asymmetry, and smoothness, see][]{conselice03} taken from \cite{ravi06} for all selected objects in our GOODS/CDF-S photometric dataset parameterized by their spectral type as in Figure~\ref{ps:spectype}. Note the strong separation of yE candidates in the concentration parameter ({\it left panel}) from galaxies with later spectral types.}
\label{ps:morph}
\end{figure*}

\subsection{Morphologies}
To investigate the yE candidate morphologies we cross-correlate the
GOODS/CDF-S sample with the structural parameter classification of
\cite{ravi06}.~Based on the $BViz$ HST/ACS images the authors computed CAS
parameters ('asymmetry', 'concentration', and 'smoothness' defined in
\citealt{conselice03}) for all our sample sources.~Because of the redshift
distribution of our sources each ACS filter probes a different wavelength
and therefore a different morphological (stellar) component in a
galaxy.~To allow a comparison of structural parameters that is independent
of redshift we compute the redshift-corrected restframe $B$-band
morphologies by linearly interpolating the CAS parameters between the
pivot wavelengths of each ACS filter \citep[see][]{sirianni05} according
to the redshift of each source.~In Figure~\ref{ps:morph} we show the
distribution of restframe $B$-band CAS parameters for all sample galaxies
sub-divided by spectral-type as in Figure~\ref{ps:spectype}.~We find that
the overall spectral-type classification of our yE candidates
is consistent with early-type morphologies, which confirms the robustness
of our procedure to select early-type post-starburst galaxies. In
particular, the distribution of the CAS 'concentration' parameter
indicates that very few yE candidate galaxies with later-type
morphologies were selected by our photometric selection.

\subsection{$BJK$ Color-Color Diagram}
\begin{figure*}[!th]
\centering 
\includegraphics[bb=0 0 400 400, width=8.9cm]{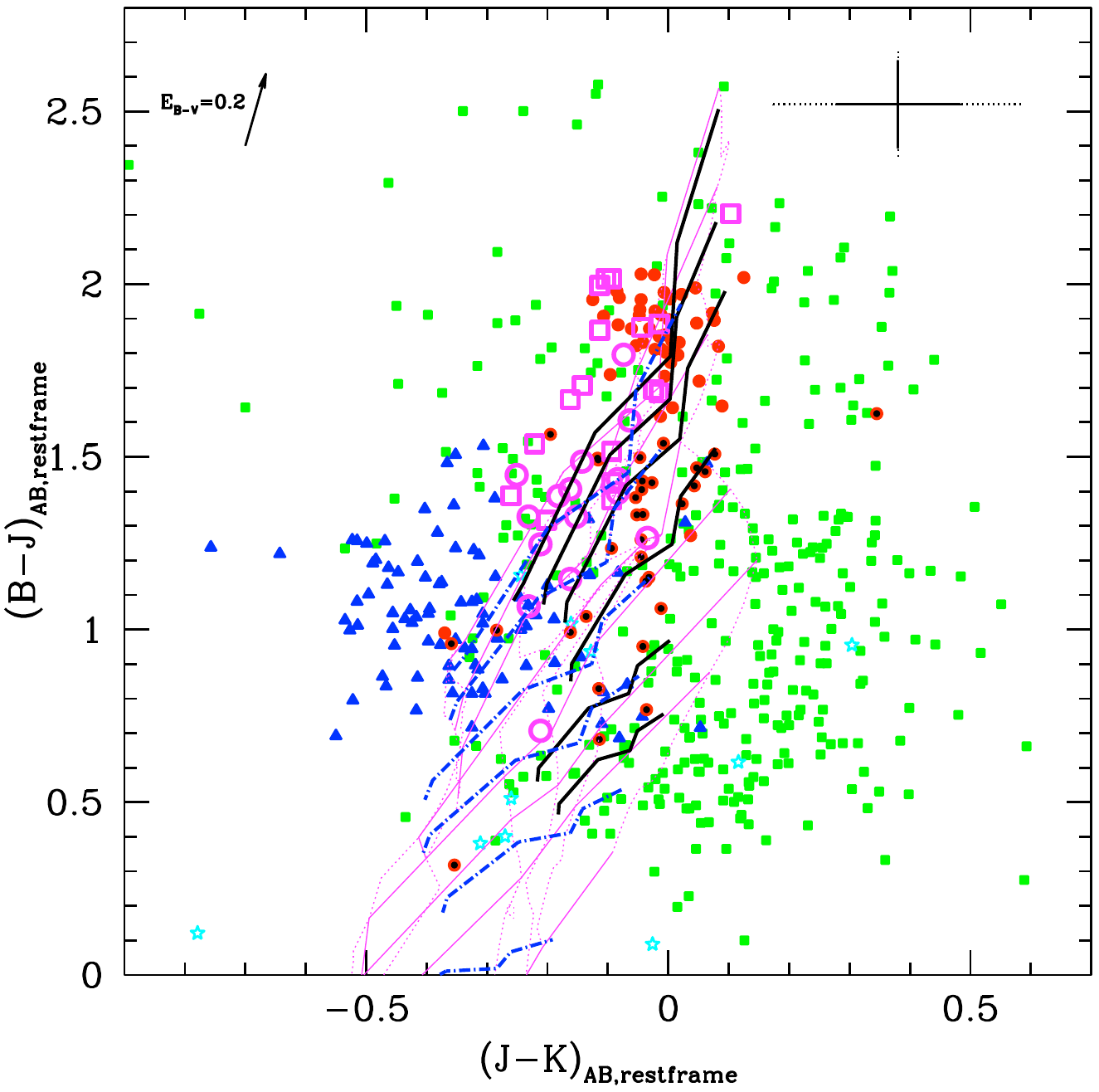}
\includegraphics[bb=0 0 400 400, width=8.9cm]{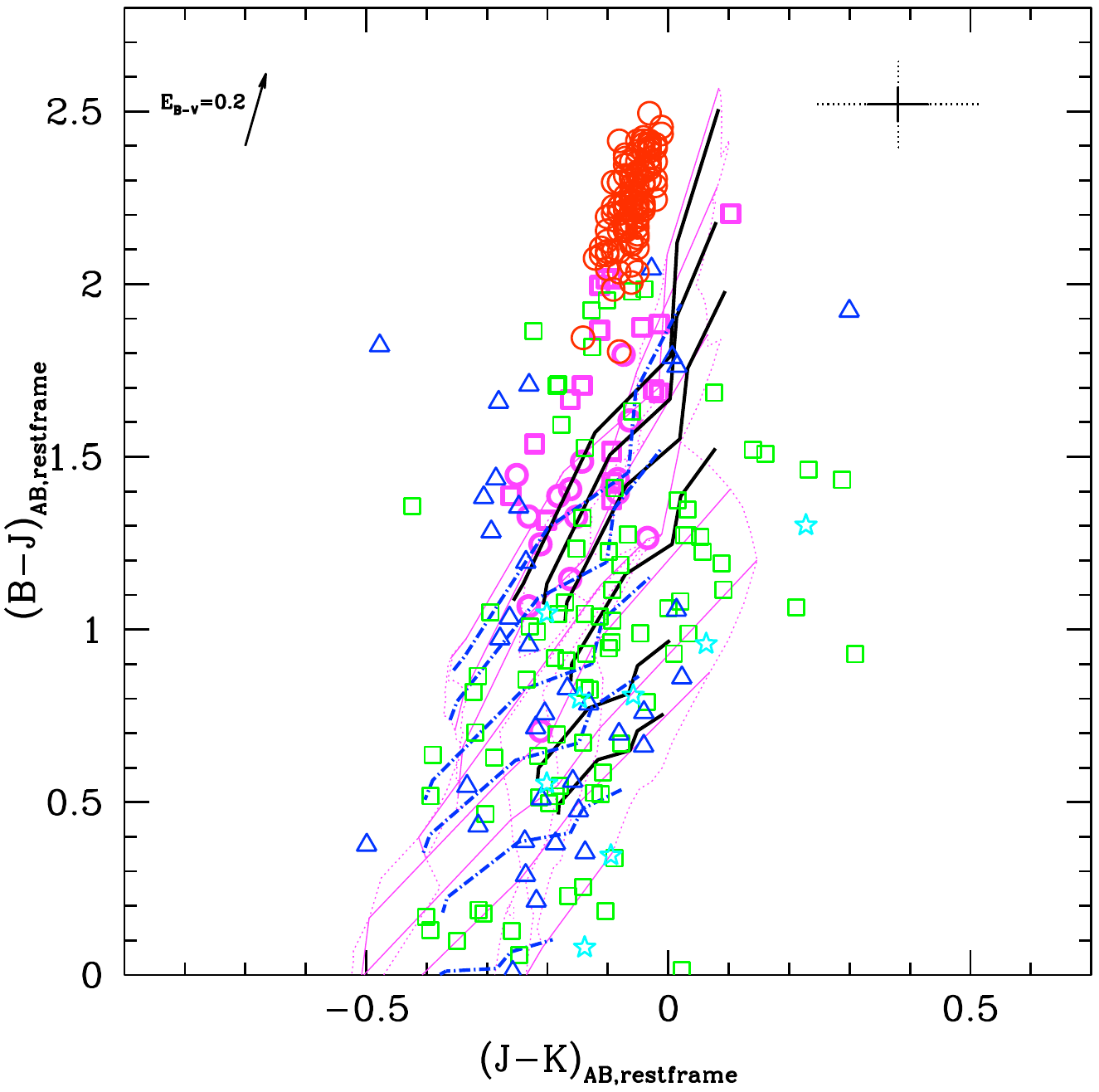}
 \caption{{\it Left panel}: $BJK$ color-color diagram (using de-reddened 
rest-frame AB magnitudes) of galaxies in the GOODS/CDF-S Field. The data
symbols are parameterized by the galaxy spectral types, and are early-type
({\it filled circles}), spiral ({\it filled squares}), irregular ({\it
filled triangles}), and starbursts ({\it stars}).~yE galaxy
candidates are marked by {\it small dark dots}. Spectroscopically
confirmed K+A galaxies taken from \cite{galaz00} are indicated by open
symbols, where open squares mark cluster K+A galaxies and open circles are
field K+As. Average total photometric errors, including uncertainties of the photometry and the photometric redshift determination, are indicated in the upper right
corner for yE candidates ({\it solid cross}) and the mean photometric uncertainties for confirmed K+A galaxies ({\it dotted cross}). {\it Right panel}: $BJK$ color-color
diagram for local galaxies. Open circles are nearby ellipticals from
\cite{michard05}, open squares and triangles are Sab and Scd spirals,
while stars show local starburst galaxies, both taken from \cite{perez03a}. The
same spectroscopically confirmed K+A sample as in the left panel is
overplotted for comparison. The solid error bar shows average
uncertainties for the elliptical sample, while the dotted cross is for the
later-type galaxies. SSP models in both panels are identical to those in
the left panel of Figure~\ref{ps:sspburst}.
}
\label{ps:bjk}
\end{figure*}

The left panel of Figure~\ref{ps:bjk} shows the GOODS/CDF-S data in the $BJK$
color-color plane. Different symbols depict different spectral types as
assigned to each object by the photometric redshift fitting routine.~Most
objects with SEDs consistent with passive early-type galaxies clump around
restframe $(B\!-\!J)_{\rm AB}\approx1.9$ and $(J\!-\!K)_{\rm AB}\approx0$.
Objects with slightly later SED types (our yE candidates) scatter towards
bluer $B\!-\!J$ colors and have intermediate $J\!-\!K$ colors that place
their mean locus between spiral and star-forming galaxies.~{\it We find
that the spectral-type selection is consistent with the $BJK$ color
selection and morphological classification.} Moreover, most yE candidates
are consistent with the previously computed composite stellar population
models for post-starburst galaxies.~Only one yE candidate at
$(B\!-\!J)_{\rm AB}\!\approx\!1.6$ and $(J\!-\!K)_{\rm AB}\!\approx\!0.35$
deviates significantly from the model predictions and might be a dusty
starburst.~For reference, we plot the control sample of spectroscopically
confirmed K+As (open symbols).
 
To provide a comparison independent of SSP model predictions we show in
the right panel of Figure~\ref{ps:bjk} nearby early-type \citep[taken
from][]{michard05} and late-type galaxies \citep[taken from][]{perez03a}
in the same color-color plane as the GOODS/CDF-S data.~We find a good
agreement between the early-type galaxies in GOODS/CDF-S and the local
galaxy  sample. However, the smaller photometric errors of the local
galaxy sample allow a more clearly detected offset of the
spectroscopically confirmed K+A galaxies towards bluer $B\!-\!J$ colors
(i.e., younger ages) compared to the local early-type galaxies, which line
up in a tight sequence. We also notice that fewer spiral (squares) and Irr
galaxies (triangles) lie outside of the SSP grid which is likely again,
partly due to smaller photometric errors of the nearby sample.~We
investigate whether this disagreement may be due to systematics in the
photometric redshift determination, but find no differences in the error
distributions between spiral and irregular galaxies, in particular in
$J\!-\!K$ that appears to be primarily responsible for the scatter.~A
matching between the photometric redshifts of our sample and currently
available spectroscopic redshifts from the literature \citep{lefevre04}
results in good agreement (see Table~\ref{photspecz}).~Although we compare
restframe colors in both samples, the galaxies in our GOODS/CDF-S data are
more distant than the nearby comparison sample. Evolutionary factors, such
as possibly enhanced dust fractions, may be responsible for the enhanced
scatter in our data. Deep mid-IR imaging and optical to infrared
spectroscopy should be able to resolve this issue.

\begin{deluxetable}{lrrr}
\tabletypesize{\scriptsize}
\tablecaption{Comparison between photometric and spectroscopic 
redshifts. \label{photspecz}}
\tablewidth{0pt}
\centering
\tablehead{
\colhead{subsample} & 
\colhead{$\langle\Delta z\rangle$\tablenotemark{a}}   & 
\colhead{$\sigma$}  & \colhead{$N$}
}
\startdata
 E    &$-0.109\pm 0.004$ & 0.061 & 14\\
 yE  &$-0.119\pm 0.012$ & 0.083 &   7\\
 Sab&$ 0.028\pm 0.004$ & 0.134 & 37\\
 Scd&$-0.061\pm 0.046$ & 0.185 &   4\\
 Starburst&$ 0.224\pm0.000$ &  \dots &   1\\
 \enddata
 \tablenotetext{a}{$\langle\Delta z\rangle=\langle z_{\rm phot} - z_{\rm spec}\rangle$.}

\end{deluxetable}

We define the population of yE galaxy candidates as those objects that
have slightly later spectral types ($T\!=\!1\onethird$, see
Fig.~\ref{ps:spectype}) and $BJK$ colors that are inconsistent with
passively evolving (reddest) early-type galaxies.~The selected yE
candidates are marked in Figure~\ref{ps:bjk} by red circles with central
black dots.~This definition might exclude some genuine K+As since the
control sample contains some spectroscopically confirmed K+A galaxies (in
galaxy clusters) with $BJK$ colors consistent with passively evolving
early-type galaxies.~Taking into account the average photometric error,
this fraction is expected to be less than $\sim\!15$\%. We point out that
even with the color, spectral-type, and morphology selection, our yE
candidate sample may be contaminated by early-type spiral galaxies that
can mimic the morphologies of elliptical galaxies (e.g., when in a largely
face-on configuration) and the SEDs of yE candidates because of their low,
but continuous star formation rates. Based on the local comparison sample
we estimate this contamination fraction between $\sim\!10\!-\!30\%$,
depending on the exact color selection. Again, deep follow-up spectroscopy
should be able to address this contamination fraction.


\section{Discussion}

\subsection{Color-Magnitude Diagrams}
Post-starburst galaxies tend to be brighter in the optical than their
passive counterparts \citep[e.g.][]{tran03}, which is likely due to their
additional $\sim\!1$ Gyr old stellar population. However, optical
photometry is not able to provide strong constraints on how the stellar
masses of these rejuvenated galaxies compare with those of passively
evolving galaxies.~The $K$-band stellar mass-to-light ratio changes by a
factor of $\sim\!3$ from $\sim\!1.5$ Gyr to 13 Gyr \citep[see
also][]{drory04}, which is about an order of magnitude smaller a change
than in the optical.~$M/L_{K}$ changes by about a factor 1.5 between 500
Myr and 1.5 Gyr (the K+A phase).~Hence, relative to optical colors a
$M_{K}$ selection defines the best approximation to a stellar-mass
selected dataset.

\begin{figure}[!ht]
\centering 
\includegraphics[bb=0 0 400 400, width=8.5cm]{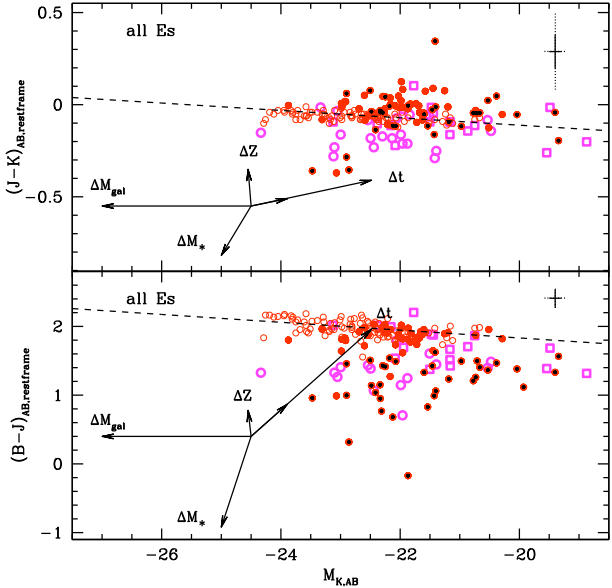}
\caption{Optical/near-infrared color-magnitude diagrams of early-type galaxies ({\it
filled circles}) and yE candidates ({\it dotted circles}).~Large open (magenta) symbols mark spectroscopically confirmed K+As \citep{galaz00} in the field ({\it circles}) and
cluster environment ({\it squares}).~Small open (red) circles are nearby elliptical galaxies from \cite{michard05}, as in the right panel of Figure~\ref{ps:bjk}. The nearby data were corrected for passive evolution, according to \cite{bruzual03} models, to the reference redshift $z=0.5$ of our GOODS data, which includes only a correction of $\Delta(B-J)=0.2$ mag.~Arrows indicate the relative changes in $\Delta t$ between 500 Myr and 1.5 Gyr (short arrow), and 500
Myr and 13 Gyr (long arrow), $\Delta Z$ between $0.2Z$ and solar
metallicity, $\Delta M_{\rm gal.}$~between $10^{10}$ and
$10^{11}M_{\odot}$, and $\Delta M_{\ast}$ between 5\% and 50\% of the
starburst mass fraction.~Color-magnitude relations for early-type galaxies from
the survey of \cite{mobasher86} are overplotted as dashed lines.~Average
photometric errors of yE candidates ({\it solid cross}) and
spectroscopically confirmed K+As ({\it dotted cross}) are plotted in the
upper right corner of each panel.}
\label{ps:cmd}
\end{figure}

In Figure~\ref{ps:cmd} we show color-magnitude (CM) diagrams for all
early-type galaxies ($T\!\leq\!1.5$) and the selected yE candidates as
well as the K+A control sample and nearby ellipticals.~Overplotted are
color-magnitude relations for early-type galaxies taken from the large
galaxy survey of \cite{mobasher86}, in good agreement with the GOODS data
and the nearby elliptical galaxy sample.~We find no significant difference
between red-sequence galaxies and yE candidates in the near-infrared
$J\!-\!K$ vs.~$M_{K}$ diagram, but we detect a significant offset in the
$B\!-\!J$ vs.~$M_{K}$ diagram where the yE candidates have significantly
bluer colors at the same $M_{K}$ (stellar mass) compared to the red
sequence.

We show the influence of changing luminosity-weighted age ($\Delta t$),
metallicity ($\Delta Z$), total galaxy mass ($\Delta M_{\rm gal.}$), and
starburst-mass fraction ($\Delta M_\ast$) as vectors in both CM diagrams.
$\Delta t$ and $\Delta M_{\rm gal.}$, and $\Delta Z$ and $\Delta M_\ast$
are highly degenerate in the $J\!-\!K$ vs.~$M_{K}$ diagram. In fact, the
$\Delta t$ and $\Delta M_{\rm gal.}$ vectors are almost parallel to the
red sequence (see upper panel in Fig.~\ref{ps:cmd}) and have, therefore,
negligible influence on the scatter in the $J\!-\!K$ color, which is
mainly driven by metallicity and the starburst-mass fraction.~The
relatively small scatter in this diagram hints at a small scatter in
metallicity and starburst-mass fraction and/or is a signature of a
correlation between these parameters.~This degeneracy can be partly lifted
with the $B\!-\!J$ vs. $M_{K}$ diagram (see below).~Four yE
candidates fall significantly off the red sequence in the $J\!-\!K$ vs.
$M_{K}$ diagram.~They are likely to have experienced significantly
different starburst events than the rest of the sample.~One possible
explanation is that the metallicity of their starburst population is
significantly lower.

The most influential parameters in the $B\!-\!J$ vs.~$M_{K}$ diagram are
age and the starburst mass fraction (besides the total galaxy mass).~Both
vectors are highly inclined relative to the red sequence, which is a
graphic explanation of the scatter in this plot.~Metallicity plays a
negligible role. We use the $B\!-\!J$ vs.~$M_{K}$ diagram to discuss the
redshift evolution of our yE candidates in the next section.

\subsection{Starburst Mass Evolution}
In the hierarchical merging scenario of galaxy formation, massive
structures are expected to form on extended timescales from smaller
sub-units with a considerable fraction of stars forming relatively
recently \citep[e.g.][]{springel05}.~If, in contrast, more massive
structures form first, as seen in the monolithic collapse scenario, the
majority of stars form at high redshifts ($z\ga2$) in intense starbursts
\citep[e.g.][]{larson75}.~The fraction of yE signatures among massive
early-type galaxies is a measure of dissipative merging activity (recently
often named ``wet merging''), which is expected to be a function of
redshift in the hierarchical merging picture. In contrast, no such
evolution is expected in the early monolithic collapse scenario out to
$z\!\sim\!2$.~This difference is expected to depend on environment and to
be more pronounced in the field \citep{benson02}, which we probe here with
the GOODS dataset.~The fraction of yE galaxies and the strength of the
yE signatures among massive early-type field galaxies as a function of
redshift therefore puts strong constraints on galaxy formation models.

\begin{figure}[!t]
\centering 
\includegraphics[bb=0 0 400 400,width=8.5cm]{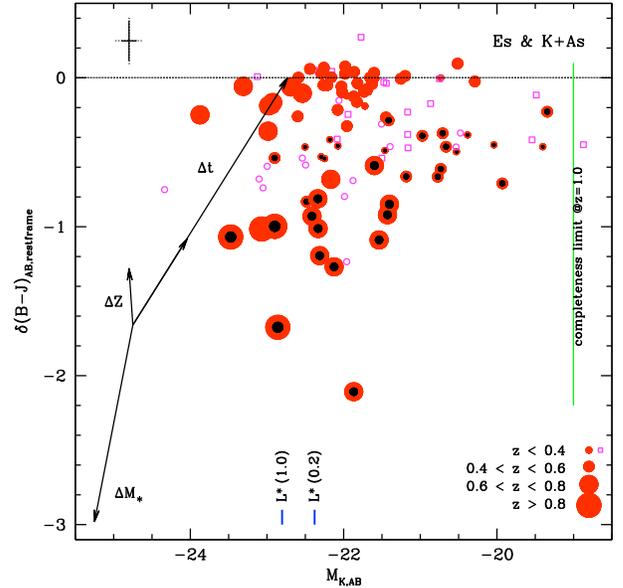}
\caption{Residual $B\!-\!J$ colors with respect to the red sequence (see
Fig.~\ref{ps:cmd}) as a function of absolute $K$ band magnitude for
early-type galaxies.~Dark dots mark yE candidates.~Open symbols indicate
spectroscopically confirmed K+A as in Fig.~\ref{ps:cmd}; taken from 
\cite{galaz00}.~The symbols size
is parametrized by redshift.~All confirmed K+A galaxies are located at
redshifts $z\leq0.12$.~Vectors indicate the influence of varying age,
metallicity, and starburst mass fraction as in Fig.~\ref{ps:cmd}.~Average
photometric errors are indicated in the upper left corner of the panel,
where the dotted cross indicates typical errors of the confirmed K+As and
the solid cross shows errors for yE candidates.~A long vertical line
shows the $10\sigma$ points source detection limit at $z\!=\!1$.~Small
vertical lines indicate $L^{\ast}$ at redshifts 0.2 and 1.0 for field
galaxies \citep{drory03}.~Note the paucity of yE candidates at magnitudes
fainter than $M_{K, {\rm AB}}\approx-21.2$ and 
$\delta(B\!-\!J)_{\rm AB}\la-0.8$ mag.}
\label{ps:bjresid}
\end{figure}

To investigate the redshift evolution of our yE candidates, we subdivide
the sample into four redshift bins between $z=0$ and 1. For each galaxy
we determine the $B\!-\!J$ offset with respect to the red sequence and
plot the residuals versus their absolute $K$-band luminosity in
Figure~\ref{ps:bjresid}.~Our analysis is limited in redshift space by the
completeness of our GOODS data in the $K$ band. A $10\sigma$ point source
detection in the VLT/ISAAC data is feasible down to $m_{K}({\rm AB})=25.1$
mag \citep{giavalisco04}, which translates into $M_{K, {\rm AB}}=-19.0$
mag at a redshift of $z=1$ (indicated by a vertical line in
Figure~\ref{ps:bjresid}).~Hence, completeness is not an issue for the
following analysis.

We observe that at any given $M_{K}$ the $B\!-\!J$ residuals are
systematically larger for higher redshift galaxies.~Spectroscopically
confirmed K+As show a similar trend, but their redshift range is
significantly smaller ($z\leq0.12$) than that of our sample ($z\leq1$, see
Fig.~\ref{ps:redshift}).~Note that confirmed field K+As show larger
$\delta(B\!-\!J)$ than confirmed cluster K+As, which lie closer to the red
sequence.~We also find tentative evidence for a systematic difference in
$\langle\delta(J\!-\!K)\rangle$ between confirmed {\it field} K+As (open
circles) and the combination of our yE candidates and confirmed {\it
cluster} K+As. It appears that a metallicity difference is the most
obvious way to explain this offset, given that in the more age-sensitive
$B\!-\!J$ panel the yE candidates lie among, or are bluer than, the nearby
confirmed field and cluster K+As.~However, the photometric errors of the
spectroscopically confirmed K+A sample require that this intriguing result
be confirmed with deep spectroscopy.

Parameters likely responsible for the evolution in $\delta(B\!-\!J)$ with
redshift are {\it (i)\/} a younger age of the post-starburst
population at higher redshift, or {\it (ii)\/} a larger mass fraction of
the post-starburst population at higher redshifts (we established in
Section~\ref{ln:ssp} that metallicity has a smaller effect on
$\delta(B\!-\!J)$). First we consider age effects. Under the assumption
that the yE candidates indeed host $0.5\!-\!1.5$ Gyr old post-starburst
populations with strong Balmer absorption-line spectra, the vectors in
Figure~\ref{ps:bjresid} show that the difference in $\delta(B\!-\!J)$
between yE candidates with $z\la0.4$ and those with $z\!>\!0.6$ is in
principle similar to the effect of age fading from 0.5 to 1.5 Gyr.
However, it is hard to imagine that post-starburst populations in yE
galaxies at higher redshift (be it induced by galaxy interactions or by
ram-pressure stripping) would be systematically younger than in
low-redshift ones, given the short duration of the yE phenomenon relative
to the difference in look-back times within this redshift range.~In the
adopted WMAP cosmology, the difference in look-back times between $z =
0.2$ and 1.0 is $7.8 - 2.4 = 5.4$ Gyr, significantly longer than the
duration of the yE phenomenon.~Instead, one would expect the ages of such
post-starburst populations to be randomly distributed (between 0.5 and 1.5
Gyr) at any redshift.~We, therefore, suggest that it is more likely that
the increase of yE candidates' $\delta(B\!-\!J)$ with redshift is
primarily due to an {\it increasing mass fraction of the post-starburst
population}.~If indeed the yE phenomenon identifies an important era in
the assembly history of early-type galaxies, the lack of low-luminosity
($M_K\!\ga\!-21$) yE candidates at $z > 0.6$ in our GOODS sample would
constitute strong evidence in favor of galaxy formation scenarios in which
more massive early-type galaxies in the field are assembled earlier than
their low-mass counterparts, which is in line with the ''downsizing''
picture \citep[e.g.][]{cowie96}.

On the other hand, we currently cannot exclude the possibility that the
higher-redshift yE candidates do not actually host a (one)
post-starburst population, but instead have systematically younger overall
(luminosity-weighted) ages than the lower-redshift yE
candidates.~More frequent bursts at higher redshifts, due to more frequent
mergers, might be a possible explanation, which would lead to
systematically younger observed ages.~Given the relevance of this issue in
terms of its power to constrain early-type galaxy formation scenarios, we
suggest a spectroscopic followup to test whether our yE candidates
really do contain K+A features in their spectra.

\subsection{Evolution of the yE Fraction}
The fraction of post-starburst galaxies in nearby clusters was reported to
increase with decreasing galaxy luminosity \citep[e.g.][]{poggianti01,
smail01}.~The morphology bias prevents a direct comparison between the
yE fractions in the cluster and field environment.~We, therefore,
select only early-type galaxies from our sample and investigate the 
yE fraction in the field.~A similar trend of an increasing post-starburst
fraction with decreasing galaxy luminosity is present in our GOODS sample
of field galaxies.~We quantify this trend by splitting the data at
$L^{\ast}/2$ along the pure luminosity evolution vector and counting the
number of passive Es and yE candidates.~The corresponding values for
the division for each redshift bin into super-$L^{\ast}/2$ and
sub-$L^{\ast}/2$ systems are $M^{\ast}_{K}=-22.27, -22.43, -22.64$, and
$-22.75$ mag for $z=0.3, 0.5, 0.7$, and $0.9$, respectively.~Note that the
samples are complete down to $M_B=-16.3$ mag and $M_{K}=-19.0$ mag at
$z\!=\!1$, which leaves our sample sensitive to objects with
restframe colors $B-K\la2.7$ mag, but excludes extremely reddened objects,
such as EROs \citep[e.g.][]{daddi00}. However, these galaxies are most
frequent beyond redshift unity \citep[e.g.][]{georgakakis06, simpson06}
and are expected not to be a significant part of our initial sample.~To
improve sample statistics we merge the number counts of the two upper and
two lower redshift bins, which results in a high and low-redshift sample
divided at $z=0.6$.~Galaxy number counts are given in
Table~\ref{tab:counts}.

We confirm the trend of a higher fraction of post-starbursts among
early-type galaxies toward less luminous objects.~In the high-redshift
sample we find a yE candidate fraction of 86\% among all sub-$L^{\ast}/2$
early-type galaxies, whereas only 22\% of the more luminous (massive)
objects are yE candidates. The yE fractions shrink towards lower
redshifts to 51\% for sub-$L^{\ast}/2$ systems and 11\% for their brighter
counterparts.~We suggest that these high yE fractions be confirmed 
(or refuted) using follow-up spectroscopy.

\begin{deluxetable}{crrrrcc}
\tabletypesize{\scriptsize}
\tablecaption{Number counts of E galaxies and yE candidates. \label{tab:counts}}
\tablewidth{0pt}
\centering
\tablehead{
\colhead{} & 
\multicolumn{2}{c}{sub-$L^{\ast}/2$} & 
\multicolumn{2}{c}{$>L^{\ast}/2$} & \colhead{} & \colhead{}  
\\
\colhead{redshift} & 
\colhead{$N_{\rm E}$}   & 
\colhead{$N_{\rm yE}$}   &
\colhead{$N_{\rm E}$}   & 
\colhead{$N_{\rm yE}$}   &
\colhead{$f_{<L^{\ast}/2}$\tablenotemark{a}} & 
\colhead{$f_{>L^{\ast}/2}$\tablenotemark{b}} 
}
\startdata
$z>0.6$ &  2 &12&   7 & 2 & 0.86 & 0.22 \\
$z<0.6$ &18 &19& 16 & 2 & 0.51 & 0.11 \\
\enddata
\tablenotetext{a}{Fraction $f_{<L^{\ast}/2}$ of yE galaxy candidates 
among all sub-$L^{\ast}/2$ early-type galaxies.}
\tablenotetext{b}{Fraction $f_{>L^{\ast}/2}$ of yE galaxy candidates 
among all $>\!L^{\ast}/2$ early-type galaxies.}
\end{deluxetable}


\section{Conclusions}
Based on the combination of photometric redshifts, spectral-type
classification, and optical/near-infrared colors, we have selected a
sample of young early-type galaxy candidates (yE candidates) from
the GOODS/CDF-S dataset.~Our technique relies on spectral-type fitting and
color-color/color-magnitude diagnostic diagrams and provides an efficient
method for selecting yE candidates using {\it only} photometric
information.~An analysis of CAS parameters shows that the selected
yE candidates have early morphological types, which confirms that
spectral-type and morphological-type selection are fully consistent with
each other.~The comparison of population synthesis models and observed
properties of a sample of spectroscopically confirmed K+A galaxies
provides strong circumstantial evidence that the selected candidates are
genuine field early-type post-starburst galaxies.

We study the systematics that drive colors and magnitudes of
post-starburst galaxies in diagnostic diagrams using current population
synthesis models.~Our analysis reveals evidence for a changing starburst
mass fraction with increasing redshift in the sense that more
massive/intense starbursts may be responsible for the yE signatures
of massive field early-type galaxies at higher redshifts.~Furthermore, we
find a higher yE candidate fraction in sub-$L^{\ast}/2$ early-type
galaxies compared to their more luminous counterparts.~Within the redshift
range of our data ($z\!\la\!1$) we may be witnessing evidence for enhanced
merging in the field towards higher redshifts, especially for
high-luminosity galaxies.~Similar results are obtained by studies of
luminosity-weighted mean ages of nearby early-type galaxies which indicate
that low-$L$ galaxies show a much greater scatter to younger
luminosity-weighted mean ages than high-$L$ galaxies
\citep[e.g.][]{caldwell03, perez03b, sanchez06}.~If follow-up spectroscopy
of the yE candidates identified in this paper confirms their K+A
nature, this suggests that the assembly of stars in high-$L$ early-type
field galaxies occurred earlier than in their lower-$L$ counterparts.


\acknowledgments

We thank Peter M. Pessev for his help in providing relevant near-IR filter
passband information. We are grateful to the referee, James Rose, for his
insightful comments that improved the content of this paper.~THP
gratefully acknowledges support in form of a Plaskett Fellowship at the
Herzberg Institute of Astrophysics of the National Research Council of
Canada. He also acknowledges support through an ESA Research Fellowship
and partial financial support through grant GO-10129 from the Space
Telescope Science Institute, which is operated by AURA, Inc.,~under NASA
Contract NAS5-26555.~This publication makes use of data products from the
Two Micron All Sky Survey, which is a joint project of the University of
Massachusetts and the Infrared Processing and Analysis Center/California
Institute of Technology, funded by the National Aeronautics and Space
Administration and the National Science Foundation.

\end{document}